# AI-washing: The Asymmetric Effects of Its Two Types on Consumer Moral Judgments

**Greg Nyilasy, PhD**
**Associate Professor**
**University of Melbourne**
gnyilasy@unimelb.edu.au

**Harsha Gangadharbatla, PhD**
**Professor**
**University of North Carolina at Chapel Hill**
gharsha@unc.edu

## ABSTRACT

As AI hype continues to grow, organizations face pressure to broadcast or downplay purported AI initiatives – even when contrary to truth. This paper introduces "AI-washing" as overstating ("deceptive boasting") or understating ("deceptive denial") a company's real AI usage. A 2×2 experiment (N = 401) examines how these false claims affect consumer attitudes and purchase intentions. Results reveal a pronounced asymmetry: deceptive denial evokes more negative moral judgments than honest negation, while deceptive boasting has no effects. We show that perceived betrayal mediates these outcomes. By clarifying how AI-washing erodes trust, the study highlights clear ethical implications for policymakers, marketers, and researchers striving for transparency.

**Introduction**

As enthusiasm for AI swells, it is understandable that organizations feel pressure to align themselves with this technological frontier. Yet, inflating or understating one's actual reliance on AI raises serious ethical questions when the 'talk' (public claims of AI use) outstrips the 'walk' (the actual reality). This practice – what we will call 'AI-washing,' inspired by the concept of greenwashing (Nyilasy et al., 2014) – challenges moral principles about transparency and trust (Lutkevich, 2024).

Consider the following two cases. In September 2023, Coca-Cola claimed to use AI to "co-create" a new product called Coca-Cola Y3000 Zero Sugar without providing any real explanation or evidence of how AI was involved or used in the process (Marr, 2024). In the second case, healthcare insurance companies such as UnitedHealth denied the use of AI and algorithms in their coverage and claims decisions while acquiring companies like naviHealth known for their predictive AI products such as nHPredict (Schreiber, 2025). In both cases it seems plausible that claims about AI use or AI non-use might have diverged from reality. Whereas the first case may overstate AI use, the other downplays or understates it.

We are interested in understanding consumer assessments of these possible deviations from the truth, and in determining which of the two deviations are perceived as a bigger betrayal compared to the respective truths. Admittedly, studying consumer responses to AI use claims, particularly, in the context of ethical consumerism or ethical consumer behavior, is a complicated problem where consumers, too, do not often walk their talk (Carrington et al., 2010). However, it is important to investigate consumer perceptions of this phenomenon as brands seem to increasingly include AI use in their messaging. This is not surprising given that research suggests that startups that mention "AI" in their messaging attract 15 to 50% more investment compared to those that do not (Marr, 2024; Olson, 2019). Using another type of managerial intuition, hiding the fact that AI was used may seem advantageous, considering algorithm aversion (Burton et al., 2020), which at the extreme may lead to tragic events such as the killing of UnitedHealthcare CEO Brian Thomson by a suspect enraged by AI-driven claim denials (Hay, 2024). These incentives make possible consumer harm from deception – in either direction – all too likely. The study of consumer perceptions and reactions to such deception is crucial to inform business ethics dialogues on the subject as well as the design of possible policy interventions.

Indeed, the Federal Trade Commission (FTC) is becoming increasingly concerned about false and unsubstantiated claims about the role of AI in advertising and asks the following questions in determining if a brand is out of bounds with their claims of AI use: Is the brand exaggerating what their AI product can do? Is the brand promising that their AI product does something better than a non-AI product? Is the brand aware of all the risk associated with their AI product? And lastly, does the product actually use AI at all? (Atleson, 2023). Apart from getting in trouble with FTC, these questions also raise serious ethical concerns for businesses and their stakeholders. AI-washing is a significant issue for business ethics because it can not only mislead and/or deceive consumers, but it can also erode trust in both the brands and the industry itself leading to unfavorable attitudes and purchase intentions.

We formally define AI-washing as *deceptive corporate practices/behaviors where organizations either overstate or understate their AI-based operations and solutions with the intention to either promote their products and services or to stave off scrutiny or criticism from key stakeholders (investors, customers, media, government, others)*. In this current study, we are interested in consumer reception of both types of AI-washing—claims overstating (*deceptive*

*boasting*) and denying the use of AI (*deceptive denial*) in their operations. In other words, the current study examines how consumers respond to AI-washing claims and how the knowledge of such deceptive practices influences their attitudes and behavioral intentions. More specifically, we are interested in testing the interaction effects of brands claiming to use AI (Says AI: yes or no) and actual use of AI in their business (Does AI: yes or no) on consumer attitudes and purchase intentions.

The contribution of our study is the introduction of the AI-washing concept into the business ethics literature. We systematically study its overstating and understating forms in an experimental setting, capturing consumer outcomes (perceptions of betrayal and overall evaluations).

In the following section, we review the very sparse mentions of AI-washing in the literature and outline the two types of AI-washing we use in our study (DB, DD). Next, we posit hypotheses about the asymmetry of the interaction effect and explain the link between the two types and consumer responses (i.e., attitudes and purchase intentions) through a mediator in the variable perceived betrayal upon learning about deceptive boasting and deceptive denial conditions. Using an online experiment, we collect data to test our hypotheses and discuss some managerial and policy implications at the end.

**Conceptual Development and Hypotheses**

*Research on the Phenomenon*

Research on AI-washing is very limited, particularly as it pertains to behavioral business ethics, with few studies examining the prevalence of this deceptive practice and the reasons why companies engage in it (Haddi, 2024). One study examined the use of AI-washing in software labeling and found that consumers are less trusting of AI-labeled systems and services (Leffrang & Müller, 2023). AI-washing has been compared to the practice of ethics washing where companies engage in superficial ethical practices without any real or meaningful commitment or action in order to stave off critics and avoid any regulation (Bietti, 2021). Another study that examined AI-washing in the banking sector argues that AI-washing could undermine consumer confidence in the long run (Prisznyák, 2023).

More broadly, the ethics of AI has been examined in a variety of contexts and situations in organizations, particularly when it comes to employee hiring and behaviors. For instance, Figueroa-Armijos et al. (2023) examined the ethics of using AI in hiring and how perceptions of using AI in the hiring process influences individuals' trust in organizations. Another study in the area of employees' behavior in organizations found that employees adhere less to unethical instructions from an AI than from a human supervisor (Lanz et al., 2024). Lastly, Yan et al. (2024) found from a series of experiments that individuals are averse to algorithm-based HR decision-making because they found it to be deontologically problematic irrespective of the quality of the decisions. Other than these studies, AI-washing has not been fully investigated in business ethics as it pertains to consumer receptions of this practice, i.e., attitudes and behavioral intentions.

In the following section, we build our theoretical expectations about the effects of AI-washing on consumers' moral judgments and formulate the hypotheses we will test in our empirical study.

*Moral Judgements and the Effects of AI-washing on Company Attitudes and Purchase Intentions*

The current understanding and use of the term AI-washing is limited to exaggerated claims of AI use. However, given algorithm aversion, it is also reasonable to assume that companies may also downplay or understate their use of AI in their products and services. Therefore, we need to distinguish two types of AI-washing:

- *Deceptive boasting (DB)*: the dishonest practice of overstating or exaggerating AI use when companies do not in fact use AI in their products or services.
- *Deceptive denial (DD)*: the dishonest practice of understating or not stating AI use when companies in fact do use AI in their products or services.

To better understand how consumers might react to each of these two types of AI-washing messages, we turn to conceptualizations of moral judgement to build our theory of perceived AI-washing and our specific hypotheses.

Moral judgements are evaluations that consumers make as a response to a moral norm violation (Zhou et al., 2022). Moral judgements are a form of moral cognition, which includes the psychological processes that allow consumers to recognize, understand, and evaluate moral and immoral messages and behavior (Malle, 2021). Evaluative moral judgements have been measured using a wide variety of instruments ranging from attitude scales, which measure self-reported responses to moral transgressions, to physiological measures (Cannon et al., 2011). The relationship between moral judgement and behavior (or behavioral intentions) has also been widely studied in the psychology literature. Both moral judgements (measured through attitudes) and moral behaviors (often measured through intentions) are in response to moral norm violations or transgressions. People often rely on moral rules or some higher order abstract principles (e.g., "thou shall not lie") when encountered with a moral norm violation and the interaction between people's interpretation of the moral situation/violation and the moral rules/principles involved determines their moral judgement and behavior (Saltzstein, 2010).

Based on the above discussion, a moral norm can be understood as when moral situations match the individual's moral rules or principles. Moral violations occur when there is a mismatch or conflict between moral situations and individuals' expectations based on their moral principles or rules. For example, a moral norm violation occurs when there is a mismatch between what a company says and does. Therefore, the two types of AI-washing introduced earlier (DB and DD) are, in essence, violations or transgressions from two truthful, base-level moral norms, which we term *honest affirmation (HA)* and *honest negation (HN),* respectively. Honest affirmation is when, in its messaging, a company says it uses AI – and in reality, it indeed does use AI. Honest negation is a situation where a company says it does not use AI and in reality, it indeed does not use it. Deceptive boasting (DB) is a violation of the moral norm of honest affirmation (HA), and deception denial (DD) is a violation of the moral norm, honest negation (HN), respectively. Moral judgements can be used to explain consumer responses in terms of attitudes toward the company and purchase intentions to each these moral norm violations when AI-washing happens.

Given that moral judgements are strongly associated with moral norm violations, the magnitude of moral norm violation should determine the nature and magnitude of consumers' moral judgements in terms of attitudes and intentions. For instance, more drastic moral violations should result in more unfavorable attitudes toward the company and purchase intentions than violations that are perceived as minor. When it comes to AI-washing, with the growing number of brands claiming AI use in today's advertising, it can be argued that consumers have come to expect a certain level of AI-washing from most brands. However, what consumers might not be expecting is for companies to downplay their AI use in products and services. This suggests an asymmetry of the interaction effect of AI-washing messages on consumer attitudes and intentions. In other words, the moral norm violation associated with deceptive denial (against honest negation) should be larger than the moral norm violation associated with deceptive boasting (against honest affirmation), resulting in bigger differences in attitudes and intentions for the former moral violation than the latter. This leads to our first hypothesis:

> **H1:** There will be an interactive effect of AI-washing on consumer attitudes and purchase intentions such that a company's deceptive denial ("says no AI, but does AI") has a significant negative effect, while deceptive boasting ("says AI, but does no AI") does not (versus respective "honesty" controls).

*The Mechanism: Perceived Betrayal*

Turning to the mechanism underpinning this effect, when it comes to moral norm violations or transgressions, "the most important process involved is the inference as to the moral actor's intentions and to the impact of the act on individuals and the group or society (Saltzstein, 2010, p. 301). This explains the asymmetry between our two types of moral violations (DB and DD). For instance, consumers may attribute a firm's deceptive boasting (DB) to several factors such as a pressure to fit in with others in the sector, a desire to increase value in the eyes of its various stakeholders, gain competitive advantage and increase market share, and even increasing consumer and media expectations, all of which relate to the moral actor's intentions to succeed in a highly competitive environment. In some ways, these intentions that result in an increase in DB messages may be considered as part of doing business in a highly competitive free market economy leading to consumers judging DB messages more favorably than DD messages.

Deceptive denial (DD) messages, in contrast, are expected to have a negative impact on individuals and the group or society for a variety of reasons. Given the prevalence of deceptive boasting, consumers might find it harder to justify or find reasons why a firm might not want to disclose their AI use, particularly in an environment where consumers, media, and investors all expect firms to use AI. Therefore, consumers are more likely to associate hiding AI use or the act of deceptive denial as more nefarious and/or unethical in nature. When consumers attribute a firm's deceptive denial strategies to covering up problematic AI use such as in the case of health insurance companies using AI to deny claims, they are more likely to judge such practices more unfavorably than DB message strategies.

The result is that consumers will feel a greater sense of betrayal (Grégoire & Fisher, 2008) when firms engage in deceptive denial than when they engage in deceptive boasting. Given the plethora of companies producing advertising messages that claim AI use in almost every single product and service category, consumers have come to expect most, if not all, companies to claim AI use in their products or services. However, when a firm understates or claims of no AI use, that should result in a bigger magnitude of moral norm violation, thereby resulting in feelings of betrayal and distrust. This brings us to our second hypothesis formulating our expectations for the mediating mechanism:

> **H2:** Consumer betrayal mediates the interactive effective between the AI-washing and the resulting consumer attitudes and intentions in H1.

**Method**

*Research Design*

To test our predictions, we designed a 2 × 2 between subjects experiment, following Nyilasy et al. (2014). Because we wanted to test the asymmetric effects of two types of AI-washing, we allowed for the full factorial combination of whether a firm says they have used AI ("Says AI") and whether they in fact have used AI ("Does AI"). Type I AI-washing (deceptive denial, a false positive) occurs when a company claims they have used AI when in fact they haven't. Type II AI-washing (deceptive boasting, a false negative) occurs when a company claims they have not used AI but in fact they have. Table 1 depicts the basic study design.

[PLACE TABLE 1 ABOUT HERE]

*Participants and Procedures*

We recruited a representative US sample from Prolific Academic. Four hundred and one respondents completed the study. Respondents gave their informed consent after they were told that the purpose of the study was to understand how they feel about the use of AI by brands and companies. In accordance with the Institutional Review Board (IRB) approval for the project, subjects were participating voluntarily, free to withdraw from the project and retained anonymity.

*Manipulation and Dependent Measures*

We manipulated two independent factors orthogonally using scenarios about a fictitious company, PWXL, a health insurer. According to the scenarios, consumers encounter this provider during their search for health insurance. In all the four scenarios identically, it becomes obvious that the respondents' original expectations are violated and there is less coverage than hoped for. In each of the four scenarios, corresponding to the 2 × 2 design, there is information provided about what the insurer claims (first factor, Says AI: Yes or no) and a screen later, what the insurer turns out to be actually doing (second factor, Does AI: Yes or no). The full stimulus set is provided in Table 2.

**[PLACE TABLE 2 ABOUT HERE]**

The dependent measures in the study were (adapted from Nyilasy et al., 2014): 1) *attitudes*, measured by asking "Now think about the company (PWXL Insurance Co.) and rate your overall attitudes toward the company. My attitude toward PWXL Insurance Co. is..." (7 pt semantic differential scales: Bad/good; Unfavorable/favorable; Negative/positive) and 2) purchase intentions, measured by asking "How likely is it that you would renew the insurance policy with this company next year?" (7 pt semantic differential scale: Extremely unlikely/extremely likely).

The mediator was respondent's sense of betrayal adapted from Grégoire and Fisher (2008), using seven-point Likert scales: "I felt cheated; I felt betrayed; I felt lied to; The insurance company intended to take advantage of me; The insurance company tried to abuse me."

Finally, covariates (prior attitudes toward the insurance industry; prior attitudes toward AI; three items both similar to dependent attitudes), demographics (age, gender, political philosophy) and manipulation check variables for each factor (binaries) were collected for robustness.

**Results**

*Descriptive Statistics*

We collected data for a total of N = 401 participants of which 198 (49.4%) were female, 195 (48.6%) were male and 8 non-binary or other. Participants' age ranged from 18 to 81, with a mean of 45 years of age (SD = 15.6). The distribution of subjects in each cell shows a nearly perfectly balanced design (Table 3).

**[PLACE TABLE 2 ABOUT HERE]**

We checked reliability for all scales using Cronbach's $\alpha$ – both overall for the entire sample and per each of the four cells. In all cases, reliabilities were well above the benchmark $\alpha$ = 0.70 value (Hair et al., 2019). Reliabilities, means and standard deviations for all multi-item dependent, mediating and control measures are provided in Table 4. Correlations in between these measures as well as the manipulated factors are provided in Table 5.

**[PLACE TABLES 3 AND 4 ABOUT HERE]**

*Manipulation Checks*

We tested if each of the factors in the 2 (Claim of AI use: "Says AI": Yes vs. no) × 2 (Actual AI use: "Does AI": Yes vs. no) design worked as intended – both in terms of their respective main effects on the manipulation check variables (MCSays and MCDoes), and also, if one factor may have confounded the effects of the other. To achieve this and knowing that the manipulation check variables were categorical, we ran two polynomial logistic regressions.

In the first, we examine whether Says AI, Does AI, and their interaction significantly predicted the likelihood that participants endorsed the statement (MCSays = Yes). The dependent variable was the transformed MCSays (0 = "No," 1 = "Yes"), and the predictors were Says AI, Does AI, and the Says AI × Does AI interaction term (Equation 1):

$$logit(p_s) = ln(\frac{p_s}{1-p_s}) = b_0 + b_1(SaysIV) + b_2(DoesIV) + b_3(SaysIV * DoesIV) \quad (1)$$

where $p_s$ is the probability of respondents endorsing "yes" when asked about the MCSays manipulation check; $b_0$ is the intercept (the log-odds of endorsing "yes" when all predictors are 0) and $b_1, b_2, b_3$ are unstandardized regression coefficients for Says AI, Does AI and their interaction, respectively.

Results showed a significant main effect of Says AI, $b_1$ = 5.31, $SE$ = 1.16, $z$ = 4.59, $p$ < .001, with an odds ratio of 202.85 (95% CI [23.38, 2281.09]), indicating that participants were far more likely to correctly identify that the firm claimed AI use when the scenario suggested this. Neither Does AI ($p$ = .73) nor the Says AI × Does AI interaction ($p$ =.52) was statistically significant. These findings suggest that the Says AI manipulation was successful and was not confounded by the Does AI factor.

We tested the second factor similarly, whether Says AI, Does AI, and their interaction significantly predicted the likelihood that participants endorsed the statement (MCDoes = Yes). The dependent variable was the transformed MCDoes variable (0 = "No," 1 = "Yes"), and the predictors again were Says AI, Does AI, and the Says AI × Does AI interaction term (Equation 2):

$$logit(p_d) = ln(\frac{p_d}{1-p_d}) = b_0 + b_1(SaysIV) + b_2(DoesIV) + b_3(SaysIV * DoesIV) \qquad (2)$$

where $p_s$ is the probability of respondents endorsing "yes" when asked about the MCDoes manipulation check, and the structure of coefficients are mirroring Equation 1.

Examination of the contingency table indicated perfect separation in one cell (all respondents correctly identified that the scenario suggested AI use when Says AI = –1 and Does AI = 1). Consequently, the standard logistic regression did not converge to meaningful parameter estimates (e.g., extremely large coefficients and confidence intervals). Because of this perfect separation, Firth's penalized logistic regression and simple follow-up binomial tests were conducted to obtain stable estimates. The overall Firth model was significant, likelihood ratio $\chi^2(3) = 390.37$, $p < .001$. Examination of individual predictors indicated that Does AI was a highly significant predictor, $b_2 = 9.83$, $SE = 2.98$, $p < .001$, with an odds ratio of 18,544 (95% CI [239, 3.26×10$^8$]). By contrast, neither Says AI ($b_1 = 2.26$, $p = .15$), nor the Does AI × Says AI interaction ($b_3 = -2.25$, $p = .08$) was statistically significant. Follow-up binomial tests assessed whether participants recognized the AI's action ("Does AI: Yes" vs. "Does AI: No") above chance levels in each of the four study cells. In the conditions where actual AI use was "No" and participants therefore were expected to respond "No," 91 of 100 (91%, $p < .001$) answered correctly in the "Does AI: No; Says AI: Yes" cell and 92 of 100 (91%, $p < .001$) answered correctly in the "Does AI: No; Says AI: No" cell. In the conditions where actual AI use was "Yes," 96 of 100 (100%, $p < .001$) recognized this in the in the "Does AI: Yes; Says AI: Yes" cell and 100 of 100 (100%, $p < .001$) did in the "Does AI: Yes; Says AI: No" cell.

These findings provide strong evidence that the Does AI manipulation was successful and was not confounded by the Says AI factor

*Testing Hypothesis 1*

Our first hypothesis suggests than there is an interaction between companies' claims of using AI and their actual use of AI (when consumers have full information about this) on consumer evaluations. The direction of this interaction is such that deceptive denial (Says no AI, but Does AI) has a significant negative effect, while deceptive boasting (Says AI, Does no AI) does not.

First, two-way factorial ANOVAs were conducted to examine the effects of whether a firm says it uses AI (Says AI) and whether it does use AI (Does AI) on consumers' attitudes and purchase intentions (PI). Results indicated a significant main effect of Says AI, $F(1, 397) = 25.81$, $p < .001$, and a significant main effect of Does AI, $F(1, 397) = 46.80$, $p < .001$. Importantly, the interaction between Says AI and Does AI was also significant, $F(1, 397) = 25.08$, $p < .001$, suggesting that the impact of stating AI use on consumer attitudes depends on whether the firm actually uses AI, in line with our expectations. Turning to PI, similarly, the results showed a significant main effect of Says AI, $F(1, 397) = 17.50$, $p < .001$, and a significant main effect of Does AI, $F(1, 397) = 30.56$, $p < .001$. Importantly, the interaction between Says AI and Does AI was significant, $F(1, 397) = 10.70$, $p = .001$, suggesting that the impact of one factor on purchase intentions depends on the level of the other, confirming our expectations.

Second, to ascertain the specific direction of the predicted interaction, we ran planned contrasts. Two specific contrasts were tested with both attitude and PI as DVs. With regards to attitudes, comparing the deceptive boasting condition ("Says AI: Yes – Does AI: No," $M = 2.44$) to the honest affirmation condition ("Says AI: Yes – Does AI: Yes," $M = 2.49$) did not yield a statistically significant difference, $t(397) = -0.25$, $p = .80$, $d = -0.04$. However, comparing the deceptive denial condition ("Says AI: No – Does AI: Yes," $M = 2.12$) to the honest negation condition ("Says AI: No – Does AI: No," $M = 3.39$) revealed a significant effect, $t(397) = 6.84$, $p < .001$, corresponding to a large effect size, $d = 0.97$. Turning to PI, comparing the deceptive boasting condition ("Says AI: Yes – Does AI: No," $M = 2.22$) to the honest affirmation condition ("Says AI: Yes – Does AI: Yes," $M = 2.04$) did not yield a statistically significant difference, $t(397) = 0.89$, $p = .37$, $d = 0.13$. However, as predicted, comparing the deceptive denial condition ("Says AI: No – Does AI: Yes," $M = 1.95$) to the honest negation condition ("Says AI: No – Does AI: No," $M = 3.06$) revealed a significant effect, $t(397) = 5.53$, $p < .001$, corresponding again to a large effect size, $d = 0.78$. Figure 1 summarizes these findings.

**[PLACE FIGURE 1 ABOUT HERE]**

*Testing Hypothesis 2*

Our second hypothesis made predictions about the mechanism underlying the interactive effect. First, we investigated the main effect of the manipulated conditions and their interaction on the mediator variable, betrayal. Levene's test did not raise a concern regarding unequal variances ($F(3, 397) = 1.60$, $p = .189$). A 2 × 2 factorial ANOVA revealed a significant main effect of Says AI, $F(1, 397) = 19.04$, $p < .001$, and a significant main effect of Does AI, $F(1, 397) = 35.89$, $p < .001$. Importantly, the interaction between Says AI and Does AI was also significant, $F(1, 397) = 22.60$, $p < .001$, suggesting that the impact of stating AI use on feelings of betrayal depends on whether the firm actually uses AI, consistent with our expectations. Planned contrasts showed the shape of this interaction. Comparing the deceptive boasting condition ("Says AI: Yes – Does AI: No," $M = 5.20$) to the honest affirmation condition ("Says

AI: Yes – Does AI: Yes," $M$ = 5.06) did not yield a statistically significant difference, $t(397)$ = 0.74, $p$ = .46, $d$ = 0.10. In contrast, comparing the deceptive denial condition ("Says AI: No – Does AI: Yes," $M$ = 5.51) to the honest negation condition ("Says AI: No – Does AI: No," $M$ = 4.37) revealed a significant effect, $t(397)$ = −5.99, $p$ < .001, corresponding to a large effect size, $d$ = −0.85, indicating higher feelings of betrayal when AI usage was concealed.

**[PLACE FIGURE 2 ABOUT HERE]**

Second, for testing the moderated mediation effect with attitudes as DV, we used PROCESS Model 8 with 5,000 bootstrap resamples and a fixed seed (Version 4.3.1 for R; Hayes, 2022). Results revealed that the use of AI negatively influenced attitude through perceived betrayal only when firm said they did not use it, $b$ = – 0.32, BootSE = 0.07, 95% CI [–0.46, – 0.20], but not when they said they did, $b$ = 0.04, BootSE = 0.05, 95% CI [–0.06, 0.14]. Thus, the index of moderated mediation was significant (IMM = 0.36, BootSE = 0.08, 95% CI [0.21, 0.54], indicating that betrayal as a mechanism explains the observed interactive effect saying and doing on consumer attitudes. Repeating this analysis, with purchase intentions, yields very similar results: the use of AI negatively influenced purchase intentions through perceived betrayal only when firm said they did not use it, $b$ = – 0.30, BootSE = 0.07, 95% CI [– 0.44, – 0.18], but not when they said they did, $b$ = 0.04, BootSE = 0.05, 95% CI [–0.06, 0.13]. Thus, the index of moderated mediation was significant (IMM = 0.34, BootSE = 0.08, 95% CI [0.18, 0.51], indicating that betrayal as a mechanism explains the observed interactive effect saying and doing on consumer attitudes. The full path models are provided in Figure 3.

*Robustness*

To ensure robustness, we conducted a number of additional tests. First, with regards to H1, we found evidence that variances were not equal across the test cells through Levene's test. Results indicated a marginal violation of the homogeneity assumption for attitudes ($F(3, 397) = 2.37$, $p = .070$) and a significant one for PI ($F(3, 397) = 5.43$, $p = .001$), suggesting heteroskedasticity. Therefore, we fit generalized least squares (GLS) models allowing each combination of Says AI and Does AI to have its own variance structure (for both attitudes and PI as DVs). For attitudes, results from a robust omnibus ANOVA (Type III) indicated that the main effect of Says AI, $\chi^2(1) = 23.26$, $p < .001$, and the main effect of Does AI, $\chi^2(1) = 43.39$, $p < .001$, were both statistically significant. More importantly, the interaction of Says AI and Does AI remained significant, $\chi^2(1) = 25.11$, $p < .001$. Similarly for PI, results showed both the main effect of Says AI, $\chi^2(1) = 14.80$, $p < .001$, and the main effect of Does AI, $\chi^2(1) = 28.41$, $p < .001$, were statistically significant – and so was their interaction, $\chi^2(1) = 10.72$, $p = .001$. These findings suggest that our evidence supporting H1 above is robust despite heteroskedasticity in the data.

Second, we added covariates to check if after controlling for them our effects remain the same. Only two of the covariates were correlated with DV (prior attitudes toward the insurance industry; prior attitudes toward AI; see Figure 5). These were entered as covariates to retest the hypotheses.

Beginning with the covariate of *prior attitudes toward the insurance industry*, results again indicated that the main effects of Says AI and Does AI on attitudes remained each significant,

$F(1, 396) = 26.41, p < .001$ and $F(1, 396) = 46.79, p < .001$, respectively. Importantly, so did the crucial Says AI × Does AI interaction, $F(1, 396) = 27.70, p < .001$. Turning to purchase intentions (PI), the main effects of Says AI, $F(1, 396) = 17.08, p < .001$, and Does AI, $F(1, 396) = 29.37, p < .001$, were also significant, as was their interaction, $F(1, 396) = 10.95, p = .001$.

Repeating the analyses with *prior attitudes toward AI* as the covariate revealed a similar pattern. For attitudes, there were significant main effects of Says AI, $F(1, 396) = 27.03, p < .001$, and Does AI, $F(1, 396) = 47.27, p < .001$, as well as a significant Says AI × Does AI interaction, $F(1, 396) = 26.99, p < .001$. Similarly, for purchase intentions, both the main effects of Says AI, $F(1, 396) = 18.09, p < .001$, and Does AI, $F(1, 396) = 30.41, p < .001$, were significant, as was their interaction, $F(1, 396) = 11.47, p = .001$.

Note that in each of the above models, the three-way interaction term was non-significant. For instance, when controlling for prior attitudes toward the insurance industry, the three-way interaction for purchase intentions was $F(1, 393) = 1.07, p = .30$, and for attitudes, $F(1, 393) = 0.22, p = .64$. Likewise, when controlling for prior attitudes toward AI, neither the three-way interaction for purchase intentions ($F(1, 393) = 0.05, p = .83$) nor for attitudes ($F(1, 393) = 0.27, p = .60$) was significant. These results confirm that the homogeneity of regression slopes assumption remained tenable in our analyses, indicating that the influence of our manipulated factors (Says AI and Does AI) did not depend on participants' baseline attitudes toward the insurance industry or AI.

**Discussion**

AI is revolutionizing every aspect of our lives, but as with any new technology, it is hard to separate the reality from hype. Companies can engage in two types of AI-washing by either overstating (deceptive boasting) or understating (deceptive denial) their real AI usage. Consumer responses to both these types of AI-washing is not the same – suggesting an asymmetry in reactions. Our findings indicate that deceptive denial evokes more negative moral judgments than honest negation (the baseline for deceptive denial), while deceptive boasting has no effects. The perceived betrayal associated with each type of AI-washing explains these outcomes. In other words, consumers appear to be less betrayed with companies overstating their AI usage than with companies hiding or understating their AI use. This could be due to the oversaturation of AI claims in advertising messages that consumers perhaps have begun to expect brands to overstate their AI use to appear attractive to stakeholders, gain competitive advantage, and increase market share. In such environments where consumers increasingly expect brands to use AI, if a company downplays or hides its AI use, it might be associated with something nefarious leading to more negative moral judgements (i.e., unfavorable attitudes and purchase intentions).

*Implications*

Our study presents numerous implications for ethics researchers, firms and brands engaging in AI-washing, and policy makers. First, from a theoretical perspective, our study suggests that AI-washing is a complex phenomenon, and its scope and definition needs to evolve and expand as the use and role of AI in business changes. Currently, the few ethics commentators who noticed AI-washing are focused on the consequences of overstating AI use. Our study challenges this narrow understanding of AI-washing by incorporate consumer moral judgements about the understating, non-disclosure and denials of AI use by firms. Consumers make moral judgements when the reality is different from the claims made in ads or messages, or in other words, when a moral norm violation has occurred. Our study supports the theoretical notion that more drastic moral violations result in more unfavorable attitudes and purchase intentions. In the case of AI use, consumers consider deceptive denials as more significant than deceptive boasting. This implies that the ethical frameworks used in future studies need to incorporate expanded definitions of transparency, consumer agency, and AI-washing.

Second, for firms and brands engaging in AI-washing, our study's findings suggest that not all messages about the use of AI are equal. When firms engage in AI-washing that overstates AI use, that might be less damaging than when firms downplay or denial AI use. This implies that firms run the risk of damaging consumer trust through increased feelings of betrayal if they try to sound more human-centric by hiding their AI use.

Lastly, the implications for policy makers and regulators start with an expansive definition and understanding of AI-washing. The current regulations and guidelines (e.g., FTC guidelines on AI-washing) focus only on the misleading claims and hype surrounding AI use, and our findings indicate that denials, downplaying, and understating AI use should also be treated as form of consumer deception. For example, regulators should consider requiring firms to include statements around AI use in messages, particularly in industries such as insurance, finance and banking, healthcare, and employment where consumer AI harm may be severe. Governmental agencies and nonprofits should actively create and revise AI literacy programs that educate consumers on where AI is typically used and why transparency matters. These AI literacy and education campaigns should incorporate not just the consequences of false claims and exaggerations of AI use but also the detrimental effects of failure to disclose or underplay AI use on consumers' rights and wellbeing.

*Limitations and Future Research*

Despite the exploratory nature of our study and its interesting findings, there are some limitations that we hope future studies can address. First, our study was conducted for a company/brand in the healthcare insurance sector, which may be perceived as especially prone to AI harm. Reassuringly, including the covariates of individuals' existing attitudes toward either the healthcare insurance industry or AI did not alter our findings in any significant way. This suggests that our results should be applicable to other industries and irrespective of prior beliefs. Nonetheless, future studies should replicate our experiment in other product and service categories. Second, our experiment was designed to compare each moral norm violation from its base (i.e., deceptive denial from honest negation and deceptive boasting from honest affirmation), which in itself does not present any severe limitations, but future studies should incorporate other types of comparisons including but not limited to comparisons across industries, comparisons across message types and formats, and comparisons across different moral norm violations. Lastly, as with most experimental studies, our study has some limitations of artificiality of the setting, external validity issues (sample limited to U.S. and Prolific users), oversimplification of AI-washing and messages to one scenario, and measurement issues (betrayal and moral judgements are complex and sometimes difficult to measure precisely). Future studies should gather data and test our hypotheses using real world data from the field rather than an experimental setting along with expanding the sample to global markets and including cross-cultural aspects of processing and reception of AI-washing messages. Additionally, future studies should incorporate measures beyond attitudes and purchase intentions to gauge moral judgements to capture the complexity and nuanced emotional and moral reactions that result from exposure to AI-washing messages.

**Tables, figures**

# TABLE 1

## Study Design: 2 × 2 Between-Subjects Experiment

| IVs | Says AI: Yes | Says AI: No |
|---|---|---|
| **Does AI: Yes** | Honest affirmation | Type II AI-washing (Deceptive denial, a false negative) |
| **Does AI: No** | Type I AI-washing (Deceptive boasting, a false positive) | Honest negation |

**TABLE 2**

**Stimulus Materials: 2 × 2 Between-Subjects Experiment**

| IVs | Says AI: Yes | Says AI: No |
|---|---|---|
| **Does AI: Yes** | **[HONEST AFFIRMATION]**<br>[FIRST SCREEN]<br>Imagine you're shopping around for health insurance products online. You come across a company, PWXL that claims **they use artificial intelligence** in providing their services.<br>[NEXT SCREEN]<br>A few months later, you become ill. While they cover most of the healthcare costs – **they do not cover as much as you expected.** You learn that the company **used artificial intelligence** to calculate your coverage. | **[DECEPTIVE DENIAL]**<br>[FIRST SCREEN]<br>Imagine you're shopping around for health insurance products online. You come across a company, PWXL that claims **they do not use artificial intelligence** in providing their services.<br>[NEXT SCREEN]<br>A few months later, you become ill. While they cover most of the healthcare costs – **they do not cover as much as you expected.** You learn that the company **used artificial intelligence** to calculate your coverage. |
| **Does AI: No** | **[DECEPTIVE BOASTING]**<br>[FIRST SCREEN]<br>Imagine you're shopping around for health insurance products online. You come across a company, PWXL that claims **they use artificial intelligence** in providing their services.<br>[NEXT SCREEN]<br>A few months later, you become ill. While they cover most of the healthcare costs – **they do not cover as much as you expected.** You learn that the company **did not use artificial intelligence** to calculate your coverage. | **[HONEST NEGATION]**<br>[FIRST SCREEN]<br>Imagine you're shopping around for health insurance products online. You come across a company, PWXL that claims **they do not use artificial intelligence** in providing their services.<br>[NEXT SCREEN]<br>A few months later, you become ill. While they cover most of the healthcare costs – **they do not cover as much as you expected.** You learn that the company **did not use artificial intelligence** to calculate your coverage. |

# TABLE 3

# Distribution of Subjects in Study Cells

| IVs | Says AI: Yes | Says AI: No |
|---|---|---|
| **Does AI: Yes** | 101 | 100 |
| **Does AI: No** | 100 | 100 |

# TABLE 4

## Descriptive Statistics

| | Full sample | | Says AI: Yes | | | | Says AI: No | | | |
| | | | Does AI: Yes (Honest Affirmation) | | Does AI: No (Deceptive Boasting) | | Does AI: Yes (Deceptive Denial) | | Does AI: No (Honest Negation) | |
| Source | α | Mean (SD) | α | Mean (SD) | α | Mean (SD) | α | Mean (SD) | α | Mean (SD) |
| --- | --- | --- | --- | --- | --- | --- | --- | --- | --- | --- |
| *Dependent variables* | | | | | | | | | | |
| Purchase intent | - | 2.32 (1.48) | - | 2.04 (1.32) | - | 2.22 (1.41) | - | 1.95 (1.25) | - | 3.06 (1.67) |
| Attitude toward firm | .97 | 2.61 (1.39) | .97 | 2.49 (1.24) | .97 | 2.44 (1.28) | .98 | 2.12 (1.24) | .96 | 3.39 (1.48) |
| *Mediator variable* | | | | | | | | | | |
| Betrayal | .92 | 5.03 (1.40) | .89 | 5.06 (1.25) | .89 | 5.20 (1.25) | .92 | 5.51 (1.32) | .93 | 4.37 (1.52) |
| *Control variables* | | | | | | | | | | |
| Prior attitude towards insurance | .98 | 3.01 (1.52) | .97 | 2.83 (1.35) | .98 | 3.05 (1.60) | .98 | 2.96 (1.49) | .97 | 3.18 (1.62) |
| Prior attitude towards AI | .98 | 4.25 (1.57) | .99 | 4.11 (1.74) | .97 | 4.34 (1.38) | .98 | 4.22 (1.67) | .98 | 4.34 (1.50) |
| Age | - | 44.96 (15.63) | - | 45.93 (16.62) | - | 44.14 (15.57) | - | 46.25 (15.59) | - | 43.55 (14.78) |
| Political philosophy – Economic issues | - | 4.29 (1.85) | - | 4.38 (1.87) | - | 4.59 (1.75) | - | 4.00 (1.96) | - | 4.20 (1.78) |
| Political philosophy – Social issues | - | 4.71 (1.85) | - | 4.77 (1.85) | - | 5.10 (1.70) | - | 4.44 (1.97) | - | 4.52 (1.81) |

# TABLE 5

## Correlation Matrix

|  | Says AI | Does AI | Attitude | Purchase intent | Betrayal | Insurance prior | AI prior | Social politics | Economic politics |
|---|---|---|---|---|---|---|---|---|---|
| **Does AI** | .002<br>*(.96)* | * | | | | | | | |
| **Attitude** | .103<br>*(.038)* | .221<br>*(.000)* | | | | | | | |
| **Purchase intent** | .127<br>*(.011)* | .218<br>*(.000)* | .730<br>*(.000)* | | | | | | |
| **Betrayal** | -.069<br>*(.171)* | -.178<br>*(.000)* | -.618<br>*(.000)* | -.538<br>*(.000)* | | | | | |
| **Insurance prior** | .043<br>*(.393)* | .073<br>*(.142)* | .311<br>*(.000)* | .172<br>*(.001)* | -.208<br>*(.000)* | | | | |
| **AI prior** | .016<br>*(.749)* | .055<br>*(.271)* | .212<br>*(.000)* | .187<br>*(.000)* | -.165<br>*(.001)* | .288<br>*(.000)* | | | |
| **Social politics** | -.124<br>*(.013)* | .054<br>*(.277)* | -.066<br>*(.188)* | -.039<br>*(.441)* | -.066<br>*(.186)* | -.065<br>*(.191)* | -.001<br>*(.983)* | | |
| **Economic politics** | -.104<br>*(.037)* | .055<br>*(.272)* | -.028<br>*(.581)* | -.022<br>*(.658)* | -.087<br>*(.080)* | -.034<br>*(.502)* | -.017<br>*(.734)* | .807<br>*(.000)* | |
| **Age** | -.005<br>*(.926)* | -.072<br>*(.150)* | .001<br>*(.979)* | -.032<br>*(.521)* | .027<br>*(.591)* | .148<br>*(.003)* | .147<br>*(.003)* | -.077<br>*(.124)* | -.068<br>*(.175)* |

*Note: Pearson correlation coefficients are provided as the first number. Exact p-values are provided in italics and parentheses underneath.*

**FIGURE 1**
**Interaction Means Plot for Dependent Variables Attitude and Purchase Intent**

**Panel A: DV Attitudes**

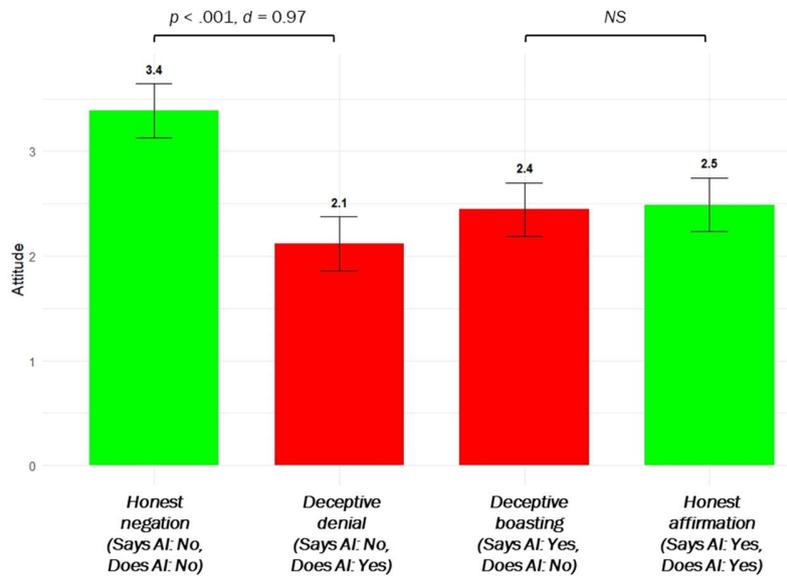

**Panel B: DV Purchase Intentions**

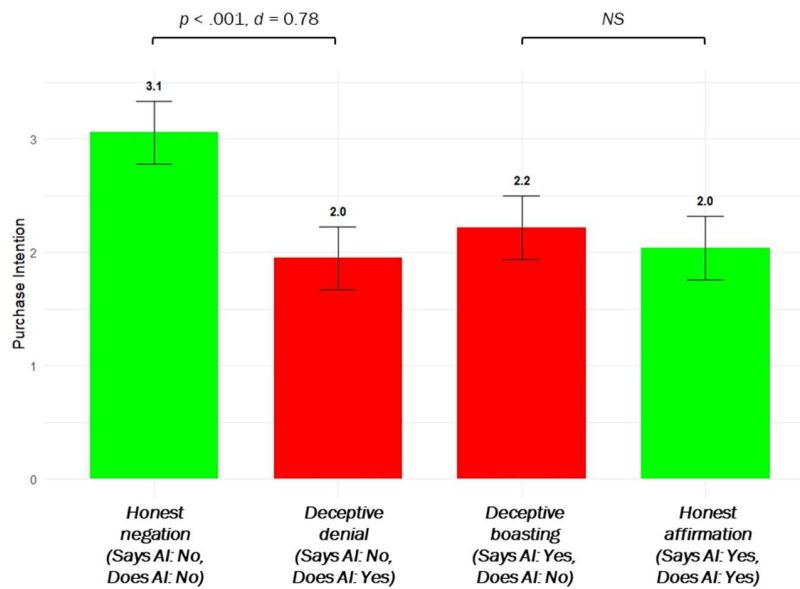

*Note: Whiskers are 95% Confidence Intervals.*

# FIGURE 2

## Interaction Means Plot for Mediator Betrayal

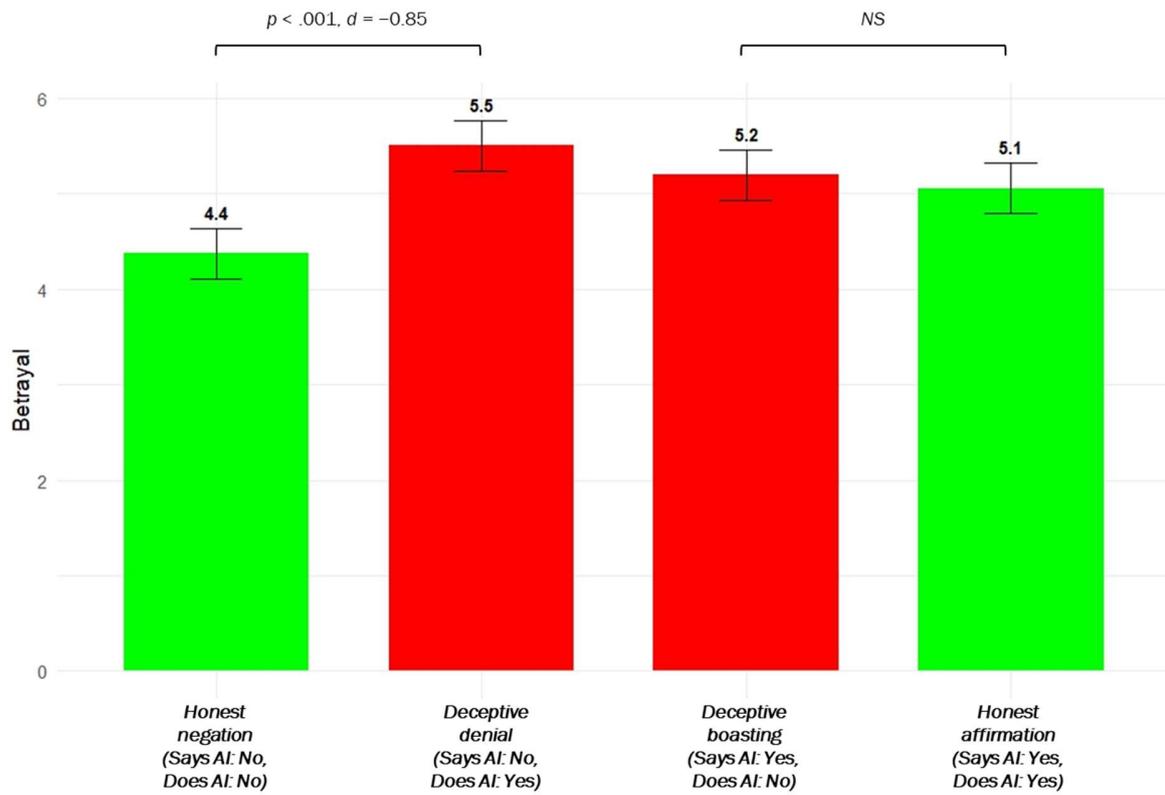

# FIGURE 3

# Moderated Mediation: Betrayal Explains the Asymmetric Effects of AI-washing Types

## Panel A: Attitudes

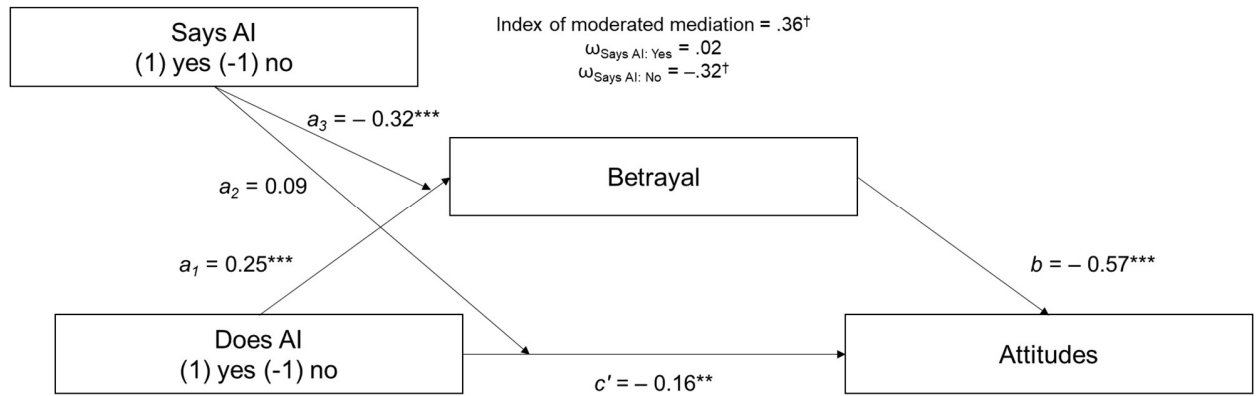

## Panel B: Purchase Intentions

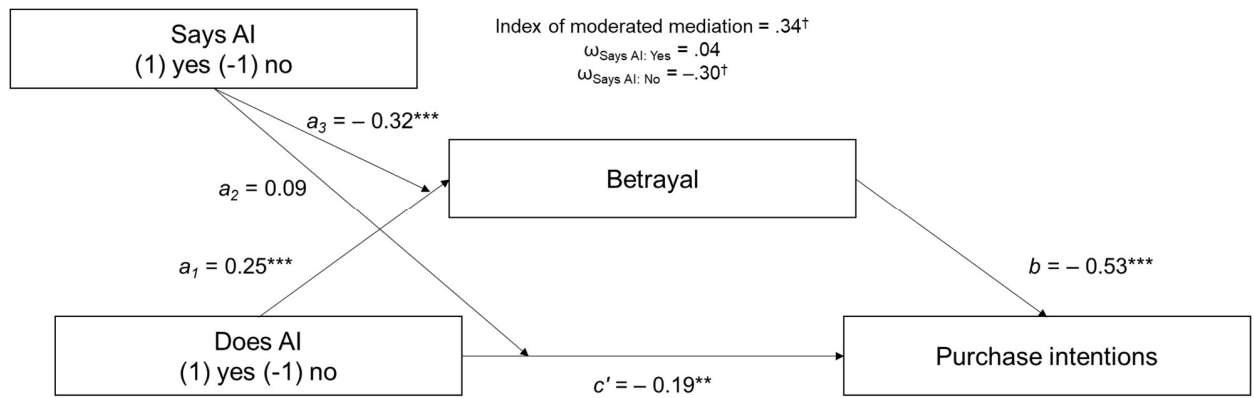

Note: *$p < .05$, **$p < .01$, ***$p < .001$; †significant at 5% level based on 95% bootstrapped confidence intervals with 5,000 repetitions; coefficients $a_1$, $a_2$, $a_3$, $b$, $c'$ are unstandardized direct effects coefficients.

**Declarations**

*IRB Approval*

This study has been approved by the Institutional Review Board (IRB) of the University of

    Colorado Boulder.

*Informed Consent*

All participants gave their informed consent prior to their inclusion in the study.

*Conflict of Interest*

The authors declare that there are no conflicts of interest.

*Open Access*